\documentclass[journal]{IEEEtran}
\usepackage[left=0.70in, right=0.70in, top=1in, bottom=1in]{geometry}
\IEEEoverridecommandlockouts
\usepackage[pdftex]{graphicx}
\usepackage[small]{caption}
\usepackage{float}
\usepackage{xspace}
\usepackage{fancyhdr}
\usepackage{amssymb,amsmath}
\usepackage{subfig}
\usepackage{xcolor,amsmath}
\usepackage{epstopdf}
\usepackage{bbding}
\usepackage{cite} 
\usepackage{color}
\usepackage{amsmath, amssymb, algorithm, algpseudocode}
\begin{document}

\title{Beyond Diagonal IRS Assisted Ultra Massive THz Systems: A Low Resolution Approach}
\author{Wali Ullah Khan, Chandan Kumar Sheemar, Zaid Abdullah, Eva Lagunas, and Symeon Chatzinotas \\Interdisciplinary Centre for Security, Reliability and Trust (SnT), University of Luxembourg\\
\{waliullah.khan,chandankumar.sheemar,zaid.abdullah,eva.lagunas,symeon.chatzinotas\}@uni.lu
}%

\markboth{IEEE International Symposium on Personal, Indoor and Mobile Radio Communications 2024}%
{Shell \MakeLowercase{\textit{et al.}}: Bare Demo of IEEEtran.cls for IEEE Journals} 

\maketitle

 
\begin{abstract}
The terahertz communications have the potential to revolutionize data transfer with unmatched speed and facilitate the development of new high-bandwidth applications. This paper studies the performance of downlink terahertz system assisted by beyond diagonal intelligent reconfigurable surface (BD-IRS). For enhanced energy efficiency and low cost, a joint precoding and BD-IRS phase shift design satisfying the $1$-bit resolution constraints to maximize the spectral efficiency is presented. The original problem is non-linear, NP-hard, and intricately coupled, and obtaining an optimal solution is challenging. To reduce the complexity, we first transform the optimization problem into two problems and then iteratively solve them to achieve an efficient solution. Numerical results demonstrate that the proposed approach for the BD-IRS assisted terahertz system significantly enhances the spectral efficiency compared to the conventional diagonal IRS assisted system.
\end{abstract}

\begin{IEEEkeywords}
Terahertz communication, beyond diagonal IRS, 1-bit resolution, spectral efficiency.

\end{IEEEkeywords}

\IEEEpeerreviewmaketitle

\section{Introduction}
Terahertz communication is an up-and-coming technology that operates within the terahertz frequency range. It boasts extensive bandwidth, high-speed data transfer capabilities, and the promise of miniaturization of components due to the small wavelength \cite{song2011present}. As conventional wireless communication channels become increasingly congested \cite{sheemar2022hybrid}, terahertz communications offer a solution to the growing demand for bandwidth-intensive applications. Despite technical challenges, advancements in materials, devices, and signal processing techniques, driving the progress of terahertz research \cite{song2021terahertz}.

One of the major challenges for the terahertz communication compared to the lower frequencies is that they suffer from severe path-loss due to molecular absorption. The ultra-massive multiple input multiple output technology combined with intelligent reconfigurable surface (IRS) has been proposed as a promising duo to overcome such a challenge \cite{10133841}. The latter can be distinguished in a fully diagonal \cite{sheemar2023irs,sheemar2023full} and beyond-diagonal \cite{li2023reconfigurable, nerini2023closed} phase response.
For the first case, Pan {\em et al.} \cite{9672716} and Hao {\em et al.} \cite{9371019} have provided spectral efficiency optimization frameworks. The work in \cite{9672716} has performed joint optimization of IRS phase shift design, subcarrier assignment, and power allocation by utilizing the block coordinate searching algorithm, while \cite{9371019} has adopted alternating optimization for hybrid beamforming and phase shift design. To this end, Wu {\em et al.} \cite{9473736} and Ren {\em et al.} \cite{9685295} have investigated the energy efficiency optimization problem in IRS assisted terahertz communication. In \cite{9473736}, a Dinkelbach method is exploited for precoding and phase shift optimization, while \cite{9685295} has adopted a distributed approach for phase shift and power control. Moreover, Qiao {\em et al.} \cite{9120206} have maximized the secrecy rate of terahertz communication by designing transmit beamforming and phase shift using an alternating optimization approach. In \cite{9840765}, Kumar {\em et al.} have investigated the bit error rate performance by passive beamforming of IRS assisted in terahertz communication. In \cite{9367288}, Pan {\em et al.} have maximized the minimum achievable rate of terahertz communication by optimizing the IRS phase shift, unmanned aerial vehicle trajectory, sub-band allocation, and power control. In \cite{10159011}, Yan {\em et al.} have studied a problem of beam split effect and beamforming design in IRS assisted terahertz communication. They maximized the acheivable sum rate by jointly optimize the IRS phase shift, time delay at transmitter and hybrid beamforming. Furthermore, in \cite{10133914}, Wang {\em et al.} have considered simultaneously transmitting and reflecting surface in terahertz communication to maximize both spectral and energy efficiency of the system by optimizing the hybrid and passive beamforming. Although several works have been proposed for the case of IRS with diagonal phase response, no contributions for the case of beyond diagonal IRS (BD-IRS) ultra massive terahertz systems are yet available in the scientific literature.

In this study, our primary objective is to address the gap in current research by introducing a novel joint beamforming design. Our focus is on optimizing both the digital beamformer and the phase response of BD-IRS to maximize the spectral efficiency. Due to an ultra-massive number of antenna elements, the integration of high-resolution digital-to-analog converters (DACs) is infeasible. As a result, we adopt a $1$-bit resolution approach for both digital beamformers and the BD-IRS. Simulation results demonstrate that our approach holds considerable promise in enhancing the spectral efficiency of the ultra-massive THz systems compared to the conventional diagonal IRS assisted configuration.

\emph{Notations:} Boldface lower and upper case case characters denote vectors and matrices, respectively. The operators $\mathbb{E}\{\cdot\}, \mbox{Tr}\{\cdot\}, (\cdot)^H$, and $\mathbf{I}$
denote expectation, trace, conjugate transpose,  and identity matrix, respectively.

\begin{figure}[!t]
\centering
\includegraphics [width=0.48\textwidth]{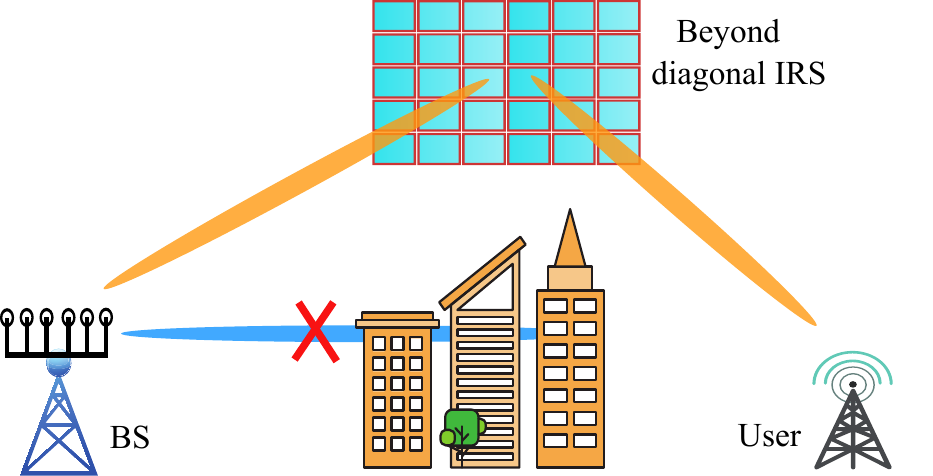}
\caption{Beyond diagonal IRS assisted THz communication, where we consider ${\bf H}\in\mathcal C^{M\times N}$ is the channel matrix between BS and BD-IRS, ${\bf g}\in\mathcal C^{M\times 1}$ is the channel vector between BD-IRS and user, and ${\bf v}\in\mathcal C^{N\times 1}$ is the effective transmitted signal.}
\label{SM1}
\end{figure}

\section{System and Channel Models}
We consider a downlink multiple input single output (MISO) THz system where a source (BS) is equipped with $N$ antennas and serves one single antenna user \cite{9743437,10133914}, as illustrated in Fig. \ref{SM1}. For high energy efficiency and low cost, we assume $1$-bit resolution DACs at both BS and BD-IRS, which have an ultra massive number of elements. As the user has only one antenna element, full resolution analog-to-digital converter (ADC) is assumed. Since terahertz is prone to high path loss and penetration loss, it is assumed that the direct link between BS and the user is blocked due to obstacles. To enhance the channel gain and ensure the line-of-sight (LoS) connectivity, a BD-IRS with $M$ reflecting element is mounted strategically to assist the signal delivery from BS to user \cite{9913356}. The BS consists of a large-scale uniform linear array (ULA), and each antenna element in the ULA is equipped with its own RF chain \cite{10159011}. Let us describe the sets of $N$ and $M$ as $\mathcal N=\{n|1,2,3,\dots,N\}$ and $\mathcal M=\{m|1,2,3,\dots,M\}$, respectively. The power gain of scattering paths in THz communication are much lower than the LoS path, hence we consider only LoS channel \cite{9677030}. Let us denote the channel matrix from BS to BD-IRS as ${\bf H}\in\mathbb C^{M\times N}$ which can be defined as:
\begin{align}
{\bf H}=\rho(f,d_1){\bf a}_{M}(\upsilon_r){\bf a}^H_{N}(\upsilon_t),\label{1}
\end{align}
Note that $\rho(f,d_1)$ is the path loss which can be described as:
\begin{align}
\rho(f,d_1)=\frac{c}{4\pi f d_1}e^{-\frac{1}{2}\mu(f)d_1}, \label{2}
\end{align}
where $f$ is the carrier frequency, $d_1$ is the distance between BS and BD-IRS, $c$ shows the speed of light, and $\mu$ denotes the molecular absorption factor. Moreover, ${\bf a}_{M}(\upsilon_r)$ and ${\bf a}_{N}(\upsilon_t)$ are the antenna array response vectors at BS and BD-IRS, defined for the respective angle of arrival (AoA) and angle of departure (AoD), which can be stated as:
\begin{align}
{\bf a}_{M}(\upsilon_r)=(1,e^{j\pi \upsilon_r},e^{j2\pi \upsilon_r},\dots,e^{j(N_t-1)\pi \upsilon_r})^T, \label{3}
\end{align}
\begin{align}
{\bf a}_{N}(\upsilon_t)=(1,e^{j\pi \upsilon_t},e^{j2\pi \upsilon_t},\dots,e^{j(M-1)\pi \upsilon_t})^T, \label{4}
\end{align}
where $\upsilon_\iota=2d_0f\sin{\theta_\iota}/c$, where $\iota\in\{r,t\}$, $d_0$ is the antenna spacing and $\theta_\iota\in\{-\pi/2,\pi/2\}$ state the AoD and AoA. Accordingly, the channel vector between IRS and user can be denoted as ${\bf g}\in\mathbb C^{M \times 1}$, which can be expressed as:
\begin{align}
{\bf g}=\rho(f,d_2){\bf a}_{M}^H(\upsilon_{t}), \label{5}
\end{align}
where $d_2$ is the distance between BD-IRS and users while ${\bf a}_{M}(\upsilon_{t})$ is the antenna array response vector, which can be written as:
\begin{align}
{\bf a}_{M}(\upsilon_{t})=(1,e^{j\pi \upsilon_{t}},e^{j2\pi \upsilon_{t}},\dots,e^{j(M-1)\pi \upsilon_{t}})^T, \label{6}
\end{align}
where $\upsilon_{t}$ can be defined as $\upsilon_\iota$.

During the communication process, the BS applies a precoding vector ${\bf f}\in\mathcal C^{N\times1}$ to the signal vector ${\bf x}\in\mathcal C^{N\times1}$ and adopts DACs to quantize the real and imaginary parts of the precoded signal using a single bit. Here ${\bf x}$ is assumed to satisfy $\mathbb E\{{\bf xx}^H\}={\bf I}$. We consider that the quantized output fall in the complex set $\Xi$ \cite{7887699,domouchtsidis2021joint}, such as:
\begin{align}
    \Xi=\{1+1j, 1-1j, -1+1j,-1-1j\},\label{7}
\end{align}
where $j$ is the imaginary unit. The signal $y$ that user receives from BS can be expressed as:
\begin{align}
{ y}=
({\bf H}^H\boldsymbol{\Phi}{\bf g})^H{\bf v}+{\vartheta},\label{8}
\end{align}
where ${\vartheta}\sim \mathcal{CN}({ 0},\sigma^2)$ is the additive white Gaussian noise, ${\bf v}={\bf f} {x}$ denotes the effective transmitted signal with
${\bf f} \in\mathbb C^{ N \times 1}$ and {$x \in\mathbb C^{ 1 \times 1}$} denoting the digital beamformer and the data stream, respectively, and ${\bf \Phi}\in\mathbb C^{M\times M}$ denotes the phase shift response of the IRS. Considering the hardware limitation of BD-IRS, we consider the phase shifts from a finite number of discrete values. Based on this, the number of quantization bits for the phase shift of $\phi_m$ reflecting element at IRS is assumed as $\phi_m\in\zeta$. Then, the set of discrete phase shift of each reflecting element available at IRS can be described as:
\begin{align}
\zeta \overset{\Delta}{=} \{\xi, \xi.e^{j\frac{2\pi}{2^L}},\dots,\xi.e^{j\frac{2\pi}{2^L}(2^L-1)}\},
\end{align}
where $ L$ shows the resolution of each element and $\xi$ shows the amplitude. Following this, each element of IRS should follow $\boldsymbol{\Phi}(i,j)\in\zeta,\ \forall i,j$, where $(i,j)$ show element at $i$-th row and $j$-th column. In this work, we assume BD-IRS with full reflection, and as such, we set $\xi=1$. Thus, $\zeta$ can be re-expressed as $\zeta = \{1, e^{j\frac{2\pi}{2^L}},\dots, e^{j\frac{2\pi}{2^L}(2^L-1)}\}$.

\section{Problem Formulation and Proposed Solution}   
The signal to interference plus noise ratio can be denoted as $\gamma$, and is expressed as:
\begin{align}
\gamma=|({\bf H}^H\boldsymbol{\Phi}{\bf g})^H{\bf v}|^2/\sigma^2
\end{align}

For the system under consideration, our goal is to maximize its spectral efficiency by jointly designing the 1-bit precoder ${\bf v}$ and the phase response of the BD-IRS, denoted by $\boldsymbol{\Phi}$. The joint optimization problem can be formulated as follows:

\begin{alignat}{2}
\mathcal P_0:\ \ & \underset{{\bf v},\boldsymbol{\Phi}}{\text{max}}\  \log_2(1+\gamma)\label{12}\\
s.t.& 
\begin{cases}\nonumber
C1:\ {\bf v}(i,j)\in {\Xi}\\
C2:\ \boldsymbol{\Phi}(i,j)\in \zeta,\ \forall i,j,\\
C3:\ \text{tr}({\bf v}{\bf v}^H)\leq P_{tot},
\end{cases}
\end{alignat}

where constraints $C1$ and $C2$ denote the low resolution imposed by DACs and the discrete phase shift design for the BD-IRS, respectively. Additionally, $C3$ represents the total transmit power constraint. 

It can be seen that $\log( \cdot)$ in \eqref{12}, along with the coupling of the decision variables ${\bf v}$ and $\boldsymbol{\Phi}$, make the optimization nonlinear, nonconvex, and NP-hard in nature. Consequently, computational challenges arise in finding the optimal solution. To overcome this, we decomposed the original optimization problem into two subproblems, iteratively solved them, and obtained the local optimal solution. Whereas the gap between local and global optimal solutions is left aside for future work. In the following, we present the proposed solution to find an efficient solution of \eqref{12}. For this, we decouple the original optimization problem into phase shift control and beamforming vector design, as expressed below:
\subsection{Low Resolution Beamforming Design}
Under the fixed value of $\boldsymbol{\Phi}$, the optimization problem for ${\bf v}$ can be stated as:
\begin{alignat}{2}
\mathcal P_1:\ \ & \underset{{\bf v}}{\text{max}}\ \log_2(1+\gamma)\label{13}\\
s.t.&
\begin{cases}\nonumber
C1:\ {\bf v}(i,j)\in {\Xi}\\
C3:\ \text{tr}({\bf v}{\bf v}^H)\leq P_{tot},
\end{cases}
\end{alignat}
The optimization problem $\mathcal P_1$ is still computationally challenging due to the low-resolution constraint. To address this, we adopt the conjugate gradient method (CG) and initialize it by setting it to belong to the feasible set of $C1$. Following that, the gradient of the objective function with respect to ${\bf v}$ at iteration $k$ is approximated numerically as: 
\begin{equation}
\text{grad}({\bf v}_k) = \left[ \frac{f({\bf v}_{i,j} + \delta) - f({\bf v}_{i,j} - \delta)}{2\delta} \right]_{i,j},
\end{equation}
Upon computing the gradient, the conjugate direction for iteration $(k+1)$, denoted as $\mathbf{p}_{k+1}$, is updated based on the gradient information and potentially previous directions under specific conjugate gradient formulations. Next, we determine the step size $\xi_k$ through a line search method that maximizes the objective function along the direction $\mathbf{p}_k$, which can be formalized as: 
\begin{equation}
\xi_k = \arg\max_{\xi} f({\bf v}_k + \xi \mathbf{p}_k).
\end{equation}
Subsequently, the beamforming matrix can be iteratively updated as follows.
\begin{equation}
{\bf v}_{k+1} = {\bf v}_k + \xi_k \mathbf{p}_k.
\end{equation}
Next, to ensure the feasibility of the constraints $C1$, ${\bf v}_{k+1}$ is projected onto the low-resolution matrix ${\Xi}$.
The termination criteria for the proposed algorithm are set such that the following condition satisfies
$
\left| f({\bf v}_{k+1}) - f({\bf v}_k) \right| < \epsilon,
$
where $\epsilon$ is a small positive value indicating the convergence tolerance.

\begin{algorithm}\footnotesize
\label{algo}
\caption{Low Resolution Beamforming and Discrete Phase Shift Design.}
\begin{algorithmic}[1]

\State \textbf{Input:} $\boldsymbol{\Phi}$, $\mathbf{g}$, $\mathbf{H}$, $\Xi$, $\zeta$
\State \textbf{Output:} Optimized $\mathbf{v}$ and $\boldsymbol{\Phi}$

\State \textbf{Initialize:} $\mathbf{v} \in \Xi$, $\boldsymbol{\alpha}$, $\boldsymbol{\beta}$, $k = 0$
\While{not converged}
\While{not converged}
    \State // \textit{Precoding Vector Design}
    \State Compute $\text{gradV}(\mathbf{v}_k)$ using numerical gradient
    \State Determine $\mathbf{p}_{k+1}$ and $\xi_k$ via CG method
    \State Update $\mathbf{v}_{k+1} = \mathbf{v}_k + \xi_k \mathbf{p}_k$
    \State Project $\mathbf{v}_{k+1}$ onto $\Xi$
    \State Check convergence for $\mathbf{v}$; Update $k$
\EndWhile

\State // \textit{Normalization of Channel Matrices}
\State $\mathbf{\Bar{g}} = \frac{\boldsymbol{\alpha} \cdot \mathbf{g}}{\sqrt{(\boldsymbol{\alpha} \cdot \mathbf{g})^T (\boldsymbol{\alpha} \cdot \mathbf{g})}}$
\State $\mathbf{\Bar{H}} = \frac{\mathbf{H} \cdot \boldsymbol{\beta}}{\sqrt{(\mathbf{H} \cdot \boldsymbol{\beta})^T (\mathbf{H} \cdot \boldsymbol{\beta})}}$

\State // \textit{Discrete Phase Shift Design}
\State Compute $\boldsymbol{\Pi}$ using $\mathbf{\Bar{g}}$ and $\mathbf{\Bar{H}}$
\State $[\mathbf{X}, \Lambda] = \text{eig}(\boldsymbol{\Pi})$; Sort eigenvalues in $\Lambda$
\State Formulate $\mathbf{Q}$ using sorted $\Lambda$ and adjust $\mathbf{X}$
\State Compute phase adjustments $\boldsymbol{\phi}$ and form $\mathbf{D}$
\State Update $\boldsymbol{\Phi} = \exp(j \cdot \text{angle}(\mathbf{\Bar{g}} \cdot \mathbf{\Bar{H}})) \cdot \mathbf{X} \cdot \mathbf{D} \cdot \mathbf{X}^T$
\State Project $\boldsymbol{\Phi}$ to discrete phase shift matrix $\zeta$
\EndWhile
\end{algorithmic}
\end{algorithm}
\subsection{Discrete Phase shift Design}
Given the low-resolution precoding ${\bf v}$, the optimization problem $\mathcal P_0$ for the phase shift design can then be simplified as follows:

\begin{alignat}{2}
\mathcal P_2:\ \ & \underset{\boldsymbol{\Phi}}{\text{max}}\  \log_2(1+\gamma)\label{14}\\
s.t.& 
\begin{cases}\nonumber
C2:\ \boldsymbol{\Phi}(i,j)\in \zeta,\ \forall i,j,\\
\end{cases}
\end{alignat}
By solving $\mathcal P_2$, we want to find an optimal phase shift matrix $\boldsymbol{\Phi}$ that enhances signal quality by intelligently reflecting it towards the intended user. To achieve this, we define the objective function of the problem $\mathcal P_2$ as $f({\bf \Phi},\boldsymbol{\alpha},\boldsymbol{\beta})$, where $\boldsymbol{\alpha}$ and $\boldsymbol{\beta}$ represent the initial controlling parameters initialized through the line search methods for any given iteration $k$. Note that $f(.)$ depends on the channels between the BS and the BD-IRS, as well as the channels between the BD-IRS and the user. Under given $\boldsymbol{\alpha}$ and $\boldsymbol{\beta}$ at iteration \( k \), the channel matrix $\bf H$ and $\bf g$ can be normalized as:
\begin{align}
\mathbf{\Bar{H}} &= \frac{\mathbf{H} \cdot \boldsymbol{\beta}}{\sqrt{(\mathbf{H} \cdot \boldsymbol{\beta})^T (\mathbf{H} \cdot \boldsymbol{\beta})}},\\
\mathbf{\Bar{g}} &= \frac{\boldsymbol{\alpha} \cdot \mathbf{g}}{\sqrt{(\boldsymbol{\alpha} \cdot \mathbf{g})^T (\boldsymbol{\alpha} \cdot \mathbf{g})}}.
\end{align}
Next, we compute a symmetric difference matrix, denoted by $\boldsymbol{\Pi}$, based on normalized channels such as:
\begin{equation}
\boldsymbol{\Pi} = \!\!\frac{1}{2} \left( \mathbf{\Bar{g}}^T \mathbf{\Bar{g}} + \left( \mathbf{\Bar{g}}^T \mathbf{\Bar{g}} \right)^T \right)\!\! - \frac{1}{2} \left( \mathbf{\Bar{H}} \mathbf{\Bar{H}}^T\! +\! \left( \mathbf{\Bar{H}} \mathbf{\Bar{H}}^T\right)^T\right),
\end{equation}
The matrix $\boldsymbol{\Pi}$ is then decomposed into its eigenvalues and eigenvectors, forming the basis of our optimization approach. It can be expressed as $[\mathbf{X}, \Lambda] = \text{eig}(\boldsymbol{\Pi})$, where $\mathbf{X}$ encapsulates the eigenvectors and $\Lambda$ is the diagonal matrix of eigenvalues. Subsequently, the eigenvalues $\Lambda$, sorted in descending order, orchestrate the rearrangement of $\mathbf{X}$, leading to $\mathbf{X} = \text{fliplr}(\mathbf{X})$.

Next, using the eigenstructure of $\boldsymbol{\Pi}$, we synthesize a transformation matrix $\mathbf{Q}$ from an identity matrix $\text{I}_{N}$ and employ matrix $\boldsymbol{\lambda}=\text{diag}(\Lambda)$  such as $\mathbf{Q}(1,1) = \sqrt{-\frac{\lambda(N-1)}{\lambda(1)-\lambda(N-1)}}$ among others, pave the way for its refinement. Concurrently, two vectors $\boldsymbol{\gamma}_1$ and $\boldsymbol{\gamma}_2$ are initialized as null vectors and then updated based on the characteristics of $\bf Q$. For instance, $\boldsymbol{\gamma}_1(1) = \mathbf{Q}(N-1,1)$, while e tailored structure of $\mathbf{Q}$: $\boldsymbol{\gamma}_1(1) = \mathbf{Q}(N-1,1)$, with $\boldsymbol{\gamma}_1(N-1) = -\mathbf{Q}(1,1)$, with similar adjustment for $\boldsymbol{\gamma}_2$. These updates result in the final configuration of $\bf Q$, which, along with another matrix $\bf X$, is crucial in updating $\boldsymbol{\Phi}$.

The phase adjustment vector $\boldsymbol{\phi}$ is determined through the relation $-\text{transpose}(\text{angle}(\boldsymbol{\beta} \cdot \mathbf{X})) - \text{angle}(\mathbf{X}^\top \cdot \boldsymbol{\alpha})$, which is then encapsulated within $\mathbf{D} = \text{diag}(\exp(j\boldsymbol{\phi}))$. These steps results in the refined transformation matrix $\boldsymbol{\Phi} = \exp(j \cdot \text{angle}(\mathbf{\Bar{g}} \cdot \mathbf{\Bar{H}})) \cdot \mathbf{X} \cdot \mathbf{D} \cdot \mathbf{X}^\top$, which facilitates an optimized signal transformation strategy. This refined matrix is subsequently projected onto the discrete phase shift matrix, where correlation is assessed to update the values of $\boldsymbol{\alpha}$ and $\boldsymbol{\beta}$ accordingly. The details of the proposed solution is also illustrated in Algorithm \textbf{1}. 

\begin{table}[!t]\footnotesize
\centering
\caption{Simulation parameters}
\begin{tabular}{|c||c|} 
\hline 
Parameter & Value  \\
\hline\hline
Carrier frequency ($f$) & 0.1 THz \\\hline
Total power budget of BS ($P_{tot}$) & 10, 20, 30 dBm \\\hline
Number of antennas at BS ($N$) & $[64-512]$ \\\hline
Number of IRS elements ($M$) & $150$ \\\hline
Bandwidth & 1 MHz \\\hline
Area length & 10$\times10$ m  \\\hline
Noise $N_o$ & -174 dBm/Hz \\
\hline 
\end{tabular}
\label{Tab}
\end{table}
\section{Numerical Results}
This section provides numerical results and discussion to evaluate the performance of the proposed BD-IRS assisted ultra-massive terahertz communication. For a fair comparison, we consider diagonal IRS (denoted as D-IRS) assisted terahertz communication as a benchmark framework. We compare the performance of proposed BD-IRS and D-IRS using different system parameters such as antenna elements and transmit power. Unless mentioned otherwise, the simulation parameters are given in Table \ref{Tab}.
\begin{figure}[!t]
\centering
\includegraphics [width=0.49\textwidth]{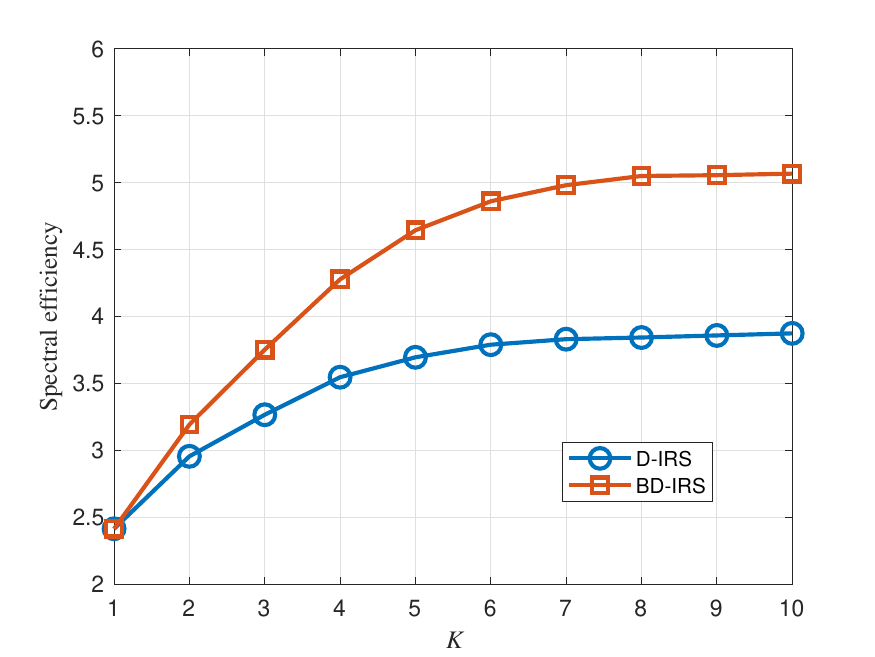}
\caption{Convergence of the proposed algorithm, for $N=M=150$, $P_{tot}=10$ dBm.}
\label{Fig1}
\end{figure}
\begin{figure}[!t]
\centering
\includegraphics [width=0.49\textwidth]{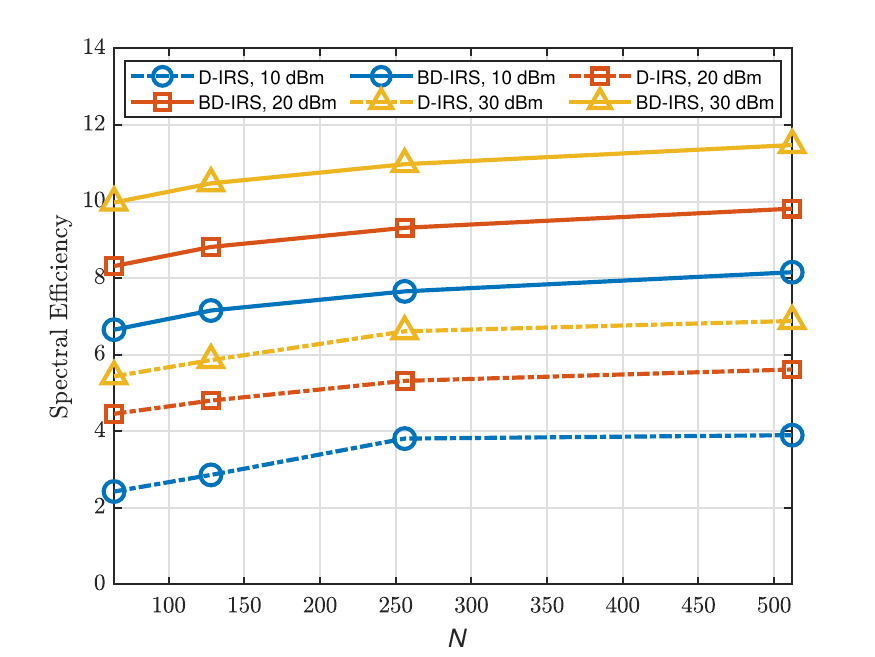}
\caption{Number of antenna elements versus spectral efficiency of the system with different transmit power levels.}
\label{Fig2}
\end{figure}
First, we provide the convergence of the proposed BD-IRS assisted terahertz system and the benchmark D-IRS assisted terahertz system optimization frameworks, which is determined by the number of $K$ iterations, as illustrated in Fig. \ref{Fig1}. For this result, we set the number of antenna elements at BS as $N=150$, the number of reflecting elements for both BD-IRS and D-IRS as $M=150$, and the transmit power at BS as $P_{tot}=10$ dBm. It can be observed that the spectral efficiency of both optimization frameworks increases with an increasing number of iterations and converges after a few iterations. It can also be noted that the proposed BD-IRS assisted ultra massive terahertz communication achieves significantly high spectral efficiency compared to the benchmark B-IRS assisted terahertz communication.

Next, we evaluate the performance of both optimization frameworks by varying the system parameters. As shown in Fig. \ref{Fig2}, the impact of antenna elements and the transmit power of BS on the system's spectral efficiency is analyzed. For this result, we set the number of reflecting elements for BD-IRS and D-IRS as $M=150$. It can be seen that increasing the number of elements and transmit power results in high spectral efficiency of the system for both BD-IRS assisted terahertz communication and D-IRS assisted terahertz communication. However, the proposed BD-IRS assisted terahertz system outperforms the D-IRS assisted terahertz system. Moreover, it can also be noticed that the increase in transmit power has a greater effect on the spectral efficiency compared to the antenna elements. In particular, an increase in transmit power contributes an approximate increase of 20\% while an increase in antenna elements exhibits an approximate increase of 15\% to the spectral efficiency of the system.

\section{Conclusion}
This paper investigated the spectral efficiency of terahertz communication by designing a low-resolution precoding at the transmitter and a discrete phase-shift design at BD-IRS. The original optimization problem was non-linear, NP hard and coupled on low resolution precoding and discrete phase shift. Therefore, obtaining an optimal solution was challenging and complex. The problem was first transformed into two problems for low resolution precoding discrete phase shift design and then iteratively solved to obtain an efficient solution. Numerical results demonstrate that the proposed terahertz communication framework with BD-IRS achieves high spectral efficiency compared to the conventional diagonal IRS assisted terahertz communication. 
\section*{Acknowledgement}
This work is funded by the Luxembourg National Research Fund (FNR) as part of the CORE program under project RISOTTI C20/IS/14773976.
\bibliographystyle{IEEEtran}
\bibliography{Wali_EE}

\end{document}